\documentstyle[twoside,fleqn,espcrc2,epsf]{article}

\title{Study of the chiral behavior of
$\langle \overline{\psi} \psi \rangle$
in quenched QCD
\author{Adrian Kaehler\address{
  Department of Physics, Columbia University, New York, NY 10027,
  USA}
\thanks{This work was done in collaboration
with Dong Chen, Ping Chen, Norman Christ, George Fleming,
Chulwoo Jung, Yubing Luo, Catalin Malureanu,
ChenZhong Sui, and Pavlos Vranas
and was supported in part by the US Department of Energy.} }
}

\begin{document}

\begin{abstract}
	The QCD phase transition is studied on $32^{3}\times8$ quenched
lattices.  One expects from simple arguments that the quenched
chiral condensate should diverge as the chiral limit is approached.
Previous studies have not been able to observe this behavior, perhaps
due to relatively large lattice spacings used.  The topic is
revisited here on $N_{t}=8$ lattices, where the physical spacing
is approximately $(2\:GeV)^{-1}$.
\end{abstract}

\maketitle

\section{INTRODUCTION}
	Quenched QCD provides an interesting opportunity to
study the role of topological effects on the axial $U(1)_{A}$ symmetry
above the chiral symmetry restoring phase transition.
Because it  will turn out that our results elucidate features
of the full dynamical theory as well, they have bearing on a
wide class of problems, in particular the order of the QCD phase
transition itself\cite{wilchek}.

	We will begin by reviewing the role of small eigenvalues
of the Dirac operator in producing a chiral condensate.  We will
then show that in the quenched approximation, where the fermion
determinant is removed from the path integral, there is
a possibility for a dramatic effect which would be seen at temperatures
just above the deconfining transition.  In this case, $U(1)_{A}$
breaking effects manifest themselves as a divergence in the condensate
because they are not regulated by suppression from
fermion loops.  This situation is particularly easy to understand in an
instanton gas picture, so we will explain it in some detail there.

	We conclude by presenting results from our own investigations
of $32^{3}\times8$ lattices on Columbia University's QCDSP parallel
supercomputer, and analyzing them in this context.

\section{THEORY}

	It is not difficult to show that the chiral condensate
is just the trace of the the Euclidean fermion propagator
($i\not\!\!D+m$).  By looking at this trace in the representation
of the eigenvectors of the Dirac operator itself, we can separate
the contributions to the chiral condensate into two parts \cite{casher}:

\begin{equation}
\:\:Tr (\frac{i\not\!\!D-m}{D^{2}-m^{2}})
= N_{\rm{zm}}\times\frac{1}{m} + \sum_{n>0} \frac{2m}{\lambda_{n}^{2}+m^{2}}
\end{equation}
	
	Here the first term represents the contribution of 
exact zero modes, $N_{\rm{zm}}$ in number,  which are 
topological in nature while the second results from the
continuum of Dirac eigenvalues.

\subsection{Instantons}

	A simple picture which will make the role of these
zero modes clear is one of dilute instantons.  Though we
will see below that more general considerations may apply,
it is particularly instructive to think for a moment in terms
of this simple concrete model.

	To first approximation, we can take {\it dilute} to
mean that there are no interactions of any kind between
widely separated instantons or anti-instantons.
In this case, the number of zero modes should go like $N_{i}$ (the
number of instantons) {\it plus} $N_{a}$ (the number of instantons). 
In such a picture, as the volume grows, so also
does the number of zero modes, and the density of exact zero
modes does not vanish in the infinite volume limit.

	We can extend these ideas to include a more plausible
situation in which there are weak interactions between the
(anti-)instantons. Even though the presence of interactions will
give a topological charge $\nu$ = $N_{i}$ {\it minus} $N_{a}$,
and thus only $|\nu|$ exact zero modes, there will still be
{\it nearly} zero modes $\lambda_{\rm{nzm}}$
shifted away from $\lambda=0$ only slightly
by the presence of the interactions.  So, although the true zero modes
in this case will be suppressed like $\frac{1}{\sqrt{V}}$ in the
infinite volume limit, the nearly zero modes will persist.  

\subsection{Expectations}

	We might not expect such configurations
to contribute significantly to an observable such as the
chiral condensate.  Even though the contribution
appears to be divergent in the chiral limit, the fermion
determinant $\det(i \not\!\!D+m)$ is suppressing the contribution
of such configurations with a factor going like $m^{N_{\rm{zm}}}$.
%
%
	In a quenched simulation however, we effectively set
$\det(i \not\!\!D+m) = 1$. This will eliminate the suppression of
such configurations, amplifying their contribution to the
chiral condensate, and thus manifesting the
$^{1}\!\!/\!_{m}$
divergence which one would na\"{\i}vely expect from (1).

	The instanton gas model leads us to
expect a finite density of instantons in enough configurations
to have weight in the path integral.  An immediate consequence
is that the quenched approximation will result in
a 1/m divergence in the chiral condensate.  Even in the
presence of weak interactions between the 
instantons, all of this will continue to hold true, so long
as the masses being studied are greater than $\lambda_{\rm{nzm}}$.

	We can see however, that the situation is based on 
a more general idea than the dilute instanton model alone.
Though such a picture gives us a concrete way of thinking about
the situation, it is possible that the $^{1}\!\!/\!_{m}$ behavior
seen in that specific approximation is really telling us something
about the general structure of the path integral itself.  Namely,
the lack of suppression of the fermion zero modes by the quark
determinant may be itself sufficient to produce this divergence.

\section{DESCRIPTION OF CALCULATION}

	Previous quenched calculations in this temperature 
region by the Columbia group\cite{shilesh}, found that chiral
symmetry remains broken when the configurations take on either
of the complex phases of the Wilson line.  Because we are ultimately
interested in an investigation relevant to unquenched
physics, we restrict our attention to the purely real phase, in which
chiral symmetry is restored above the deconfining transition.

	If we hope to resolve topological effects
such as instantons, we must insure that the lattice spacing
is sufficiently fine that these features are accurately represented.
If the features we are looking at are indeed instantons, we would
expect that the small ones would begin to decouple as a result of
asymptotic freedom.  Taking the rho Compton wavelength as
an approximate intrinsic scale of QCD, it is objects of this
size that we might expect to need to be able to resolve.
$N_{t}$=4 for $\beta\approx\beta_{c}$ gives only
$a_{\rm{lattice}} \approx(1\:GeV)^{-1}$.
Cutting the spacing in half by going to $N_{t}=8$,
gives us a lattice spacing closer to $(2\:GeV)^{-1}$, and increasing
confidence that we are probing the interesting physical region.

	In a quenched simulation, we are free to measure a 
variety of valance masses on the same ensemble of configurations
because the mass plays no role in the generation of the ensemble.
We do not however choose to measure extremely light masses, because
the lattice itself provides a cutoff for the smallest available
eigenvalues for $i\not\!\!D$. Masses below this cutoff are
in effect all identical for the purposes of computing the
chiral condensate.  For confined physics, we know that
the smallest eigenvalues have size set by the inverse total
volume. This is the basis of the appearance of nonzero
eigenvalue density around the origin and nonzero chiral condensate.  
This represents a severe compression due to QCD dynamics
relative to the free Dirac eigenvalues, for which $\lambda_{\rm{min}}$
is of order $\frac{\pi}{L}$ where L is just the extent of the
physical volume along its longest side. 
This density decreases sharply as we increase $\beta$ through the
transition, and continues to be slowly suppressed as we continue
to yet higher temperatures.  For this reason, $\frac{1}{V}$
is a lower bound on $\lambda_{\rm{min}}$.
For a volume of $32^{3}\times8$ = $2^{18}$, we expect masses
less than or equal to about $m_{q}$ = $10^{-6}$ to be uninteresting
for this reason.

	The $N_{t}$=4 Columbia study mentioned above was performed
at $\beta=5.71$\footnote{$\beta_{c_{N_{t}=4}}$ was taken to
be 5.69 \cite{christ}}.  It is clear that running at unnecessarily
high temperatures will suppress the signal we are looking to observe,
while running too close to the transition introduces the new
difficulty of tunnelings into the confined phase.
 For $N_{t}=8$, taking the critical
coupling to be $\beta_{c_{N_{t}=8}}=6.06$ \cite{karsch}, the
corresponding coupling would be
$\beta_{c_{N_{t}=8}}+\Delta\beta=6.08$. 
We chose to run here
also to make comparison with this existing $N_{t}=8$ data
as meaningful as possible.

\subsection{Simulation Details}
	This simulation was run on a 512-node prototype of
the Columbia University 8192-node QCDSP supercomputer\cite{machine}.
The prototype sustains a peak performance of 25 Gigaflops.  The 
evolutions were generated by a simple heat bath algorithm
for pure gauge SU(3) \cite{cabbibo,kennedy}. The spectrum data
represents 50 measurements, separated by 250 sweeps of the pure
gauge evolution, using a conjugate-gradient inverter to
calculate the chiral condensate with a staggered fermion Dirac
operator.  The source vectors for the CG inverter are generated
randomly to produce a noisy estimator for the fermion propagator
trace.  All masses are measured using the same fermion source
vector.  This calculation was done on a $32^{3}\times8$ volume at
$\beta=6.08$.

\section{RESULTS}
	We observe no signal for the expected $^{1}\!\!/\!_{m}$
divergence.  In fact for the mass range
$10^{-5}\leq m\leq 5\times10^{-2}$ the curve is fit very nicely
by a power law with an exponent of 0.91 (${\chi^{2}/dof}=2.0$) .
This can be compared to the $N_{t}=4$ data, where
a similar fit was found to well accommodate the data, with a power
of 0.74 (${\chi^{2}/dof}=1.9$) .
\begin{figure}[tb]
\epsfxsize=\hsize
\epsfbox{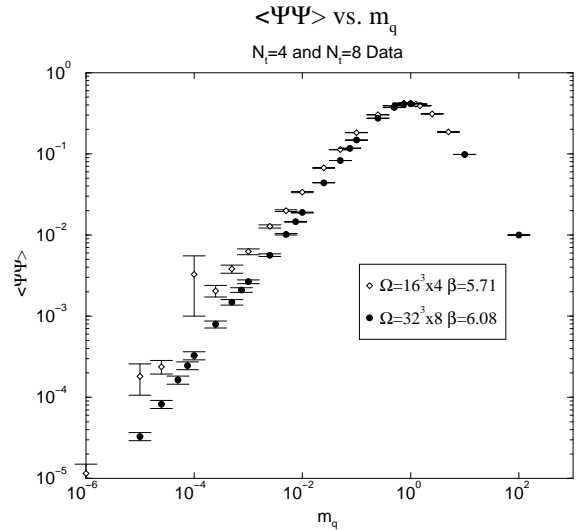}
\caption{ For $N_{t}=4$ and $N_{t}=8$, we see a power law fits 
the available data over a wide range of masses. }
\label{fig:psibarpsi48}
\end{figure}

\section{CONCLUSIONS}
	We see no evidence for the divergence predicted by the simple
instanton model.  Two possible conclusions are reasonable.  It is
possible that the simulation is in some way insufficient to probe
the physics from which this divergence derives it's origin.  On the
other hand, it is quite possible, that the opinion that there is
validity to the prediction of this divergence beyond the dilute
instanton approximation, is itself severely misplaced.

%

\end{document}